# The Neurobiology Of Thinking, Identity, And Geniality

Abstract:
Mathematically the axioms of representation are subtle, and critical. The CNC expresses its function via its internal neuronal networks in multidimensional, intrinsic frames. CNS operation can be represented by tensor transformations within and among **general frames of reference**. So-called metaorganization principle is transforming the dual representations of intention from one to another. The property of **intentionality** refers to the fact that **consciousness** is about objects or events, and is modulated by attention. To construct a new **neural framework** we must outgoing basicly from the competitive interactions among widely distributed groups of neurons. A major system in this regard is **the thalamocortical system**. The integrative dynamics of conscious experience suggests that the thalamocortical system behaves as a **functional cluster** that interacts mainly with **itself.**

## The Motor Intention as a Representation of Self-generated Action

Mental imagery is an important tool for many cognitive tasks such as perspective changes, problem solving and motor learning. With the advent of neuroimaging methods, converging evidence has been found for the involvement of area V5/human MT in mental rotation, presumably reflecting the imagination of the virtual object movement. The role of inferior and posterior parietal regions is acknowledged. Subdivisions of the intraparietal sulcus (IPS) are known to process 3-D visual information in **ego/, allo- and object-centered reference frames** and to be involved in visuo-spatial attention.

The parietal cortex has been hypothesized to evaluate imagined motor performance by comparing reafferent signals with stored internal representations of motor plans. The parietal regions might fulfil a **coupling function** between visual and motor processes.

In non-humans primates neurons in superior parietal cortex (**area 5**) code for spatial position of the contralateral arm in **body-centered coordinates,** in **the integration** of **somatosensory** and **visual information.**

The area 5 is not merely concerned with spatial position of a visual stimulus but also with its **identity.** An area 5 neurons also process somatosensory information from contralateral limbs. Along the ascending somatosensory pathway form the periphery to area S1 and to area 5, area 5 seems to be <u>the first stage</u> at which visual information about contralateral arm position is integrated with somatosensory information. The bimodal integration could form the basis for the complex **body schema,** needed for posture and to guide movement. These finding suggests a possible role for the superior parietal lobe in the **integration** of somatosensory and visual inputs about location and identity of contralateral limbs. This information is transferred to premotor areas involved in **planning** of motor acts. The motor imagery component (imaging hand movement) seems to be based primarily on motor resources under instructions to perform imagery in **the first person,** as opposed to **third person**, can be conceptualized as a comparison between reafferent sensory signals and a stored representation of the motor plan. When we imagine movements of body parts this constant monitoring of the spatial position and orientation of the imagined body part is required to achieve a **convergence** between imagined and physically executable movements. Comparing activation during executed rotational hand movements revealed a substantial overlap of neuronal responses in the posterior parietal lobe.

This is in accordance with findings that the superior parietal lobe is sensitive to both visual and somatosensory input. This also confirmed a prominent role of **the parietal cortex** for motor imagery. (5)

The study of motor functions showed that not only does the F5 area accomodate purely motor neurons, but it also houses neurons that fire when a recorded monkey **observes** another monkey, or even an experimenter, performing a similar action. These neurons were designated as **„mirror neurons".**

Subsequent experiments showed that mirror neurons are not restricted to the premotor cortex, but are also found in other areas of the monkey brain, and in homologous areas in the human brains: in the posterior parietal cortex. Mirror neurons firing also when an action is merely heard were defined **„audio-visual mirror neurons".**

Other people's intended actions we recognise by a **direct mapping** of the visual representation of the observed action onto own motor representation of the same action. The role of intentionality in triggering motor resonance phenomena is demonstrated by recent neuroimaging studies. The left primary motor cortex and parietal cortex were selectively activated when participants perceived possible human movement paths. The study confirmed that the mirror property of the premotor cortex is biologically tuned: to activate the mirror system, the action must be part of the motor repertoire of subject watching it.

On the intraindividual level, neural representations are **shared** in the sense that they are activated in different action modalities. Neural representations are shared not only by the same structures for different types of action, but also by **different brains.** When two agents interact socially with one another, the activation of **mirror networks** creates a shared neural representation, simultaneously activated in the brains of two agents. **Self-generated** actions and other people actions are mapped **onto the same** neural substratum, the same representations are activated in boths agents.

There is firm neuroimaging evidence that the interior parietal cortex and the insula are crucial componets of the mechanism involved in perceiving **the spatial features of movements.** The right posterior sulcus acivation positively correlates with temporal delay introduced **on-line** between an action and its visual feedback. In motor cognitive situation involving two people agent A generates a representation of **self-generated** action, a **motor intention.** If it is executed, it will become signal for agent B, and will form an action of the representations of the action he sees. The interaction two agents should depend on the interaction of the representations of the observed and executed **actions within each of the two brains.** Determining who is acting: **the other or the self,** should be based on **the non-overlapping part.** In **the first-person perspective** specific activation was observed in the inferior parietal lobule in **the left** hemisphere. By contrast, activation was found in a symmetrical area of the **right** hemisphere for **the third-person perspective.** The fact that a difference in activation was newertheless observed in the right inferior parietal lobe shows that it is not simply involved in associating actions and their sensory consequences, but also in **distinguishing the self from others.**

The cerebral correlates of first- and third-person perspective-taking at the conceptual level are similar to those already detected at the motor level, i. e. the right inferior parietal lobe and the frontopolar and somatosensory cortices. **The mirror system** is well designed for representing the agent's **motor intention**, but not her **prior intention** to execute an action. The prior intention is formed in advance, representing goal states. If action is a causal and **intentional transaction** between the mind and the world (consciousness), prior intention can be said to initiate a transaction, by reperesenting the end or aim of the action before the action is undertaken.

Shifting from intention-in-action to prior intention determines the mode of the intention: to **I-mode**, or **we-mode intention,** distiguishing between private aim-intention and social aim-intention (people communicating through gestures). Seeing two agents communicating resulted in activation in the medial prefrontal cortex, especially in the anterior paracingulate cortex (prospective social intentionality).

There are no significant activation differences between human and the computer conditions in the non-cooperator group. A similar activation of the anterior paracingulate cortex was found in a PET experiment involving **competitive interaction.** The anterior paracingulate cortex is activated in representing social aimes, independently from interaction type (cooperative vs. competitive), time (present vs. future), and modality (participated vs. observed).(6)

In the visual system space is encoded at the level of the retina based on the position of the activated **photoreceptors.** Visual space must initially be represesnted in an **eye-centered reference frame**. In the auditory system, spatial information must be computed based on differences in intensity and **timing** of the stimulus at the two ears and spectral cues resulting form reflections of the stimulus by the torso, head, and pinnae. Since the ears are fixed to the head, auditory spatial information should be represented in a **head-centered reference frame**. In the nervous system these two reference frames must somehow align in order to create the unified percept of a single object. A central question is what reference frames the nervous system uses to encode the spatial attributes of single- and **multimodal** stimuli and how these **reference frames** could be used to generate **unified percepts.**

For example, in the parietal lobe and the superior colliculus the responses of neurons to auditory stimuli are modulated by the eye position, so most of these neurons do not represent space in a purely head-centered reference frame. The inferior colliculus has been considered a **relay nucleus**, in which inputs from auditory brainstem converge and are then relayed to the talamus. One could expect these neurons to encode acoustic space in a head-centered reference frame and not an eye-centered reference frame. Groh et al. (9) found that approximately **one-third** of the neurons showed **neither** a head-centered **nor** an eye-centered reference frame, but **something in between.** (8) Concerning flow of information results of W. Zurek shows that the presence of redundancy divides information about the system into the three parts: classical (redundant), purely quantum, and **the borderline** (undifferentiated or „nonredundant") information. Information about the system $S$ is obtained by measuring its environment $E$. As you read this paper, you measure the albedo of the page. Actually your eyes are capturing **photons** from the electromagnetic environment. Information about the page is inferred from assumed correlations between text and photons. Researchers also gets information about $S$ by capturing and measuring of $E$.

**Quantum mutual information (QMI)** is defined in terms of the von Neumann entropy $H = -Tr(p \log p)$ as: $I_{A:B} = H_A + H_B - H_{AB}$. Unlike classical mutual information, the QMI between system A and B is **not bounded by the entropy** of either system. In the presence of **the entanglement**, the QMI can be as large as $H_A + H_B$, which reflects the existence of quantum correlations beyond the classical one. (12)

**Frame of References Perspective Transformations**
Imitation can be defined to be a **sensorimotor mapping** that transforms **sensory** information, usually visual, into corresponding **motor** commands. The vectorial expressions among general frames transformed by CNS can be summarized in a 3-step tensorial scheme to <u>transfer</u> <u>covariant sensory coordinates</u> to <u>contravariant</u> components expressed in a different **motor frame**.

**1./** <u>Sensory metric tensor</u> ($g^{pr}$), transforming a covariant reception vector ($s_r$) to contravariant perception ($s^p$). Lower and upper indeces denote co- and contravariants:

$s^p = g^{pr} s_f$ where $g^{pr} = |g_{pr}|^{-1} = |\cos(\Omega_{pr})|^{-1}$ and $|\cos(\Omega_{pr})|$ is the table of cosines of angles among the sensory unit vectors.

**2./** <u>Sensorimotor covarinat embedding tensor</u> ($c_{ip}$) transforming the sensory vector ($s^p$) into covariant motor intention vector ($m_i$). Covariant embedding is unique, including the overcompletness, results in a non-executable covarinat expression: $m_i = c_{ip} \cdot s^p$ where $c_{ip} = u_i \cdot w_p$ and $u_i$, $w_p$ are the i-th sensory unit vector and p-th motor unit vecor.

**3./** <u>Motor metric tensor converts intention</u> $m_i$ to executable contravariants, $m^e = g^{ei} \cdot m_i$ where $g^{ei}$ is computed as $g^{pr}$ for sensory axes in 1./. (4)

It requires cognitive capabilities for the recognition of conspecifics and the attribution of „others" intentions, or states of mind. The question is how one can map motions performed by **others** onto his/her **own perspective**. It requires the capacity to perform arbitrary **frames of references transformations** and to generate coherent sensorimotor mappings. In „true imitation" the imitator must be capable to extract the purpose, **the intention** of a given **sequence** of movements. To be capable of other's action **understanding** here means a goal – directed sequence of movements.

For the psychology of correlated neuronal process was important the discovery in 1992 of the **mirror neurons system,** direct-mapping mechanism between visual and motor systems. They react to highly processed stimuli represented in a **goal-centered frame of reference (FR).** The series of FR, required for transferring information in **retina-based FR** into a **body-centered FR**, is encoded by different cells **along the visual pathway**, following a sensory gradient of increasing complexity. Along the visual pathway („the what stream") the information flows from the primary visual cortex (V1) to the temporal lobes, including the inferior temporal area (IT), and the superior temporal sulcus (STS). IT contains populations of neurons that separately exhibit sensitivity to a variety of objects. These populations are sensitive to the size and orientation relative to an **viewer-centered FR,** whereas others react in an **object-centered FR.**

Spatial visual properties such as direction, orientation and size of objects are also encoded in PPC, the distance of the target objects and observed bodies, neurons activities in the ventral pathway and parietal cortex is correlate, firing differently for a stimuli in a modulatory fashion. All these regions are tightly **coupled** and form a complex network that plays a fundamental role in primates ability to reproduce movements and goal-directed actions such as transforming **viewer-centered information** into an **object-centered representation**. We try to understand the core circuits underlying the ability to **map** goal-directed motion performed by **others** into a frame of reference located onto one's **own body.**

In mathematical therms, the question is **how can we transform** a vector $\vec{v}$ given in a referential $R$ into $R'$, knowing the vector $\vec{v}_T$ across the origins of the two referentials, and the axes of the referential $R'$ itself, expressed in $R$. They are given by

$$R = \{O, \vec{e}_1, \vec{e}_2, \vec{e}_3\}, \qquad R' = \{O', \vec{e}'_1, \vec{e}'_2, \vec{e}'_3\} \quad \text{where} \quad \vec{OO'} = \vec{v}_T, \text{ and } \vec{e}'_i, \vec{e}_i,$$

$\forall_i \in \{1..3\}$ correspond to **the principal axes**, as **unit vectors**, of the demonstrator's body and observer's body, respectively. These axes correspond to the right-left, feet-head and back-front axes, respectively. The orientation of $R'$ with respect to $R$ is given by the rotation matrix: $M_{R'} = (\vec{e}'_1 | \vec{e}'_2 | \vec{e}'_3)$. We are writing down the classical transformations and considering $M_R$ as an **identity matrix**. If we consider that $M_{R'}$ is orthonormal, we know that $M_{R'}^{-1} = M_{R'}^T$. This allows us to write down using a dot product and we find: $\vec{v}' = \sum_{i \in \{1..3\}} (\vec{e}'_i \cdot (\vec{v} - \vec{v}_T)) \vec{e}'_i$. This way, the vector $\vec{v}'$ pointing to the target in **the demonstrator's**

**referential** can be directly mapped into **the imitator-centered referential**, so the demonstrator's target is considered **as the imitator's one**.

In order to extract the information from a populations of neurons was proposed to use **the population vector**. Considering that each neuron **votes** for its preferred direction proportionally to its firing activity, by taking the average of all these votes, we obtain the vector encoded by this population vector. The neurons in STS and IT are sensitive to different orientations or views of bodies and objects, respectively. These neurons activities have also been shown to be correlated to different **frames of references**, mainly in a viewer, object or goal centered reference frame. Let us consider $\Omega$, a continuos population of neurons where each unit participating in the population is characterized by its preferred directions $\vec{r}$. The preferred directions are assumed to be uniformly distributed along a 3 dimensional subspace $\Gamma = \{\vec{r} \in \Re^3 \mid \|\vec{r}\| = 1\}$, that corresponds to the surface of a unitary sphere. The response of the whole population, the population vector, is given in a continuos form by $\vec{P} = \frac{1}{K} \int_\Gamma f(u_{\vec{r}}) \vec{r} \, d\vec{r}$, where $K = \frac{2\pi}{3}$ is a normalization factor, $u_{\vec{r}}$ the neuron's membrane potential with preferred direction $\vec{r}$, and $f(u_{\vec{r}})$ its firing activity, $f$ is a non-linear function equal to $f(x) = \max(0, x)$.

Neural network perform arbitrary frames of reference transformation by applying sources of information arising from populations of neurons that encode $\vec{e'}_i, i \in \{1..3\}$, $\vec{v}$ and $\vec{v}_T$. To compute the dot product, we can use three **gain fields** whose modulatory inputs are connected to the populations coding for principal axis $\vec{e'}_i, i \in \{1..3\}$, while their vectorial inputs are linked to the difference between populations coding for $\vec{v}$ and $\vec{v}_T$ that are connected using excitatory and inhibitory **synapses**, respectively. These gain fields project to another population that will receive the result of the transformation: the vector $\vec{v}'$ in a body or object centered frame of reference using the synaptic weights $\forall_i \in \{1..3\}$.

Using **the activity profile** of the gain fields, each neuron of the final population receives a synaptic input equal to $w^{GFO_i \to \vec{v}'}_{(\vec{r},\vec{s}) \to \vec{r}^f} = \sum_{i \in \{1..3\}} ((\vec{v} - \vec{v}_T) \cdot \vec{e}^f)(\vec{r}^f \cdot \vec{e}_i) = \vec{r}^f \cdot \vec{v}^f$

This equation means that this population is encoding $\vec{v}^f$ in a **body** or **object centered frame of reference.** The visual system also provides the body and hand position in a viewer centered frame of reference, $\vec{v}_T$ and $\vec{v}$, respectively. This information in neural network compute the target location in the demonstrator´s body centered reference frame. It is directly applied to a **self-centered frame of reference** that gives the imitator its own target. To allow to reach the target, this position of the target with respect to the imitator is fed to an **inverse kinematic algorithm** that provides the sequence of joint angles. (3)

**Brain in the Brain: Glial Cells**
A series of recent papers suggests that **glia control the extent of synapse formation.** (11) Soluble factors released by macroglial cells stregthen **synaptic transmission** (retinal ganglion cells – RGCs)**.** Two follow-up studies showed that the glial factor increased the number of synapses about seven-fold. **Identity** of the synaptogenic activity depend from cholesterol, which is produced by astrocytes and secreted in apolipoprotein E – containing lipoproteins. This finding suggests that neurons require **glia-derived cholesterol** to **form** numerous and efficient synapses. How does cholesterol promote synapse formation: does it act as a synaptogenic signal, possibly after conversion to steroids, or does it serve as building material ? The massive increase in synapse number may require large amounts of cholesterol that neurons must import from astrocytes. This may explain why most synapses are formed after differentiation of astrocytes.

**Glia-derived signals regulate** the maturation of **the post-synaptic density (PSD).** Glial cells enhance the quantal size which represents the magnitude of postsynaptic responses based on enhanced postsynaptic receptor **clustering.** Glial cells increase the size of glutamate-induced whole-cell **currents** in RGCs.

Selective elimination of synapses during brain development may contribute to **structural remodeling in the adult brain** (for example in the cerebellum)**.** Experimentally induced retraction of Bergmann glia process from Purkinje cells, which had attained monosynaptic innervation, leads to **reinnervation** by multiple fibers, in a quarter of neurons.The astrocytic sheath around neurons limits the density of synaptic inputs. Glia release soluble factors, for example **proteases,** which destroy the extracellular matrix components that maintain synaptic stability. This would allow glial processes to invade the synaptic cleft and to eliminate the synapse. Taken together, the results summarized above shed new light on the synapse-glia affair.

The discovery of **two-way astrocyte-neuron communication** has demonstrated the inadequacy of the previously-held view regarding the purely supportive role for these **glial cells.** Today we can hypothesize that **glia** plays a role of „**brain in the brain**". It is now well-established that astrocytic mGluR receptors **detect** synaptic activity and respond via activation of the calcium-induced calcium release pathway, leading to elevated $Ca^{2+}$ levels (17). The spread of these levels within a **micro-domain** of one cell can coordinate the activity of disparate synapses associated with the same micro-domain. It is possible to **transmit information directly from domain to domain** and even **from astrocyte to astrocyte**, if the excitation level is strong enough to induce either intracellular or intercellular **calcium waves.** There is also information flow in the **opposite** direction, from astrocyte to synapse, via the detection of the modulation of synaptic transmission as **function of the state of the glial cells.** We must take into account both a deterministic effect of high $Ca^{2+}$ in the astocytic process, the reduction of the post-synaptic response to incoming **spikes** of the presynaptic axon, and a stochastic effect, the increase in the frequency of observed miniature **post-synaptic current** events uncorrelated with any input. There are also direct NMDA-dependent effects on the post-synaptic neuron of astrocyte-emitted factors. A **coupling** allows the **astrocyte** to act as a **gatekeeper** for the synapse. Data transmitted across the synapse can be modulated by astrocytic dynamics, which can be controlled mostly by other synapses, and the gate-keeping will depend on dynamics external to the specific synapse. The dynamics may depend on excitation from **the self-same synapse,** the behavior of the entire system is determined **self-consistently.** The spontaneous bursting activity in these networks is regulated by a set of rapidly firing neurons which we refer to as **spikers.** These neurons exhibit spiking even during long inter-burst intervals leading to some form of **self-consistent self-ectitation.** These neurons are containing astrocyte-mediated **self-synapses** (autapses).

The basic calcium phenomenology in the astrocyte arising via the glutamine-induced production of $IP_3$**.** The dynamics of the intra-astrocyte Ca level depends on the intensity of the pre-synaptic **spike train** acting as **an information integrator.** An excitatory neuron exhibits repeated spiking driven at least in part by **self-synapses.** The role of an associated astrocyte with externally imposed temporal behavior means that the astrocyte dynamics is itself determined by feedback from this particular synapse. This same modulation can in fact correlate multiple synapses connecting distinct neurons coupled to the same astrocyte. The effect of this **new multi-synaptic coupling** on **the spatio-temporal flow of information** is important. (10)

The state of composite object (consisting of the system $S$ and the environment $E$) can be ignorant of the state of $S$ alone. Entanglement between $S$ and $E$ enables envariant and implies ignorance about $S$. Envariance is associated with **phases of the Schmidt decomposition** of the state representing $SE$ (**envariance thinking algorithm**). It anticipates the consequences of environment – induced superselection („einselection") of the preferred set of pointer states, they remain unperturbed to immersion of the system in the environment. The state of combined $SE$ expressed in the Schmidt form is: $|\psi SE\rangle = \sum_{k=1}^{N} \alpha_k |\delta_k\rangle |E_k\rangle$. The properties of the entangled state $\psi SE$ do not belong to $S$ alone. The only effect on the system is the rotation of the phases of the coefficients in the Schmidt decomposition: $u_S |\delta_k\rangle = e^{iw_k^S} t_S |\delta_k\rangle = e^{i\varphi k} |\delta_k\rangle$. Phases of the coefficients in the Schmidt decomposition can be changed by local interactions. We are affecting phases of the coefficients solely by acting on the states of $E$. Schmidt states are in an intimate relationship with the pointer states as „**instantaneous pointer states**". Entangled state vector is: $|\psi SE\rangle = \sum_{k=1}^{N} |\alpha| e^{i\varphi k} |\delta_k\rangle |E_k\rangle$. (13)

**Moving Frame in the Movie of Consciousness**

In the thalamocortical system a Schrödinger picture photons are treated as an ensemble of bosons and the evolution of the many-photon probability amplitude is studied. This intuitive approach led to the great success in the quantum theory of solitons. We can obtain analytic solutions from the linear boson equations in the Schrödinger picture. The many-boson interpretation was applied in study of entangled photons. The Schrödinger picture offer an interpretation of temporal entanglement propagation. The analysis of a two-photon vector soliton consisting of two photons in orthogonal polarizations under the cross-phase modulation effect demonstrate the use of the Schrödinger picture in a temporal imaging system.

The physical significance of each amplitude $\psi_{jk}$ is that its magnitude squared gives the probability density $P_{jk}$ of coincidentally measuring one **photon** in mode $j$ at $(z,t)$ and another photon in mode $k$ at $(z',t')$: $P_{jk}(z,t,z',t') = |\psi_{jk}(z,t,z',t')|^2$. **Temporal entanglement** is defined as the irreducibility of $|\psi_{12}|^2$ into a product of one-photon ampitudes in the form of $a(t)b(t')$. **The probability** of detecting a photon in mode 1 at time $t$ is correlated to the probability of detecting a photon in mode 2 at $t'$.

The entangled properties of the photons are contained in the two-photon amplitude, the Schrödinger picture allows to use the temporal effects to engineer the entanglement. Under **the self-phase** modulation effect, cross-phase modulation offers the distinct possibility of entangling two photons in different modes. Defining time coordinates in a moving frame,

$\tau = t - \bar{\beta}_1 z$, $\tau' = t' - \bar{\beta}_1 z$, $\bar{\beta} = \dfrac{\beta_{11} + \beta_{12}}{2}$, $\Delta = \dfrac{\beta_{11} - \beta_{12}}{2}$,

than $\psi_{12}(z,\tau,\tau') = \left(\dfrac{\partial}{\partial z} + \Delta \dfrac{\partial}{\partial \tau} - \Delta \dfrac{\partial}{\partial \tau'}\right)\psi_{12} = \left[-\dfrac{i\beta_2}{2}\left(\dfrac{\partial^2}{\partial \tau^2} + \dfrac{\partial^2}{\partial \tau'^2}\right) + i\eta\delta(\tau - \tau')\right]\psi_{12}$.

This equation is a simple linear Schrödinger equation, describing a two-dimensional „wave function" $\psi_{12}(z,\tau,\tau')$ in a **moving frame** subject to a $\delta$ potential. The delta function

$\delta(t-t')$ couples the two subspaces of $\Psi_{12}(z,t,t')$, so entanglement can emerge from unentangled photons. Hence, a two-photon vector soliton generates temporal entanglement with positive time correlation as it propagates.

The $\delta$ potential enforces $S$ to take on the value $S = -\dfrac{\eta}{\beta_2}$, where $\eta$ and $\beta_2$ must have opposite signs. **The final solution** of $\psi_{12}$ in the frame of $\tau$ and $\tau'$ is therefore

$$\psi_{12}(z,\tau,\tau') = \exp\left[-i\left(\frac{\eta^2/4 + \Delta^2}{\beta_2}\right)z\right] \exp\left[-\left|\frac{\eta}{2\beta_2}\right||\tau - \tau'| + i\frac{\Delta}{\beta_2}(\tau - \tau')\right]$$

$$\int_{-\infty}^{\infty} \frac{d\Omega}{2\pi} \phi(\Omega) \exp\left[-i\Omega\left(\frac{\tau + \tau'}{2}\right) + \frac{i\beta_2}{4}\Omega^2 z\right].$$

The two-photon coherence time of a vector soliton is fixed, and **generates temporal entanglement** with positive-time correlation as it propagates. Multiple bound-state solutions can be obtained via conventional techniques of solving the linear Schrödinger equation. These general equations govern the temporal evolution of two-photon probability amplitudes in different coupled optical modes. The formalism inspires the concept of quantum temporal **imaging**, which can manipulate **the temporal entanglement of photons.** (1)

Schrödinger dynamics could lead to to a **subjective resolution („thinking")** of the measurement problem, to neuronal emergence of a **new percept**, idea in our consciousness. We perceive this **resolution** as **the definite states** in sensory bases rather than superpositions of neuronal states.

The observation involves the interception of photons that have interacted with object of interest and whose status is entangled with the state of the subject. These photons then contain indirect and redundantly coded information about the object . The conclusion of existence of superpositions of brain states is inescapable. The brain with $10^{11}$ neurons should not be thought as a deterministic classical computer with a predictable input/output pattern, since **synaptic transmission** a fairly high failure rate due to the complexity of the human brain with $10^{14}$ synapses. If each neuron is firing in average several times per second, it leads to a high degree of **unpredictability** on the everyday level.

Due to **entanglement the combined object-photon system** is described by a superposition of the form $|\psi_{oP}\rangle = \dfrac{1}{\sqrt{2}}(|\omega_1\rangle_o |\phi_1\rangle_P + |\omega_2\rangle_o |\phi_2\rangle_P)$, where $\omega_i, i=1,2$, are the two distinct spatial regions associated with object, and $|\phi_i\rangle_P$ denote the corresponding classically distinct **collective photon states**. Initial detection of such a collection of photons in the human eye is associated with rodopsin molecules in retina. As long-term memory is based on structural changes in the brain's neuroplasticity, due **the formation of new connections of synapses** between neurons and due to internal changes in the synaptic regions in individual neurons. The states in **superposition** of neuronal firing patterns will rapidly **entangle** with orthogonal (macroscopically distinguishable) states of the environment (macroscopic global superposition). (14)

**The resolution** from many coding flexibilities and streaming performance is possible due to **reference picture selection (RPS),** where each **predicted frame** can **choose** among a number of frames for motion estimation („**thinking**", associations, pictures). Consider an $M$ - frame sequence with one intra-coded frame ($I$ − **frame**) followed by $M-1$ inter-coded frames ($P$ − **frames).** Denote the frames by $F_i$, $i \in \{1,...,M\}$. We model the decoding dependencies of the sequence using a directed acyclic graph ($DAG$) $G = (V, E)$ with vertex set $V, |V| = M$, and edge set $E$. Each frame $F_i$, represented by a node $i \in V$, has a set of outgoing edges $e_{i,j} \in E$ to nodes $j$'s. $F_i$ can use $F_j$ as reference if $\exists e_{i,j} \in E$. We define $x_{i,j}$ to be the binary variable indicating wheter $F_i$ uses $F_j$ as reference. Equivalently, we define $x_{i,j}$ given $i$: $x_{i,j} = \begin{cases} 1 & if F_j asRF \forall_j \in \{v e_{i,j} \in E\} \\ 0 \end{cases}$ otherwise. Since a $P-$frame can have only one reference frame, we have the $RF$ constraint $\sum_{\{j/e_{i,j} \in E\}} x_{i,j=1} \quad \forall_i \in V$, $i \neq 1$.

We assume that only frames in the past are used for reference, e. i. $\forall e_{i,j} \in E$, $i \leq j$. It is impractical to use a reference frame too far in the past, we limit the number of candidate reference frames for any given predicted frame $F_i$ to $E_{max} << |v|$. Only frame 1 is intra-coded, and hence $\beta e_{1,j} \in E$. Associated with edge $e_{i,j}$, we denote by $r_{i,j} \in I$ the integer number of bytes in frame $i$ when frame $j$ is used as reference. The byte size of starting is $r_{1,1}$. The size $F_i$ depends on chosen $F_j$, as well as RF for $F_j$ and so on, $r_{i,j}$ is an approximate. We assume a set of QoS levels $Q = \{0,1,...,Q\}$ is available for all transmissions paths. QoS level $q_i$, transmission path $t_i$ and frame size $r_{i,j}$ will induce a **frame delivery success probability** $p_{t_i}(q_i, r_{i,j}) \in \Re$, where $0 \leq p_{t_i}(q_i, r_{i,j}) \leq 1$. Generally, $p()$ depends on $r_{i,j}$ because a large frame size will likely negatively impact **the delivery success probability of entire frame („effective thinking"). (7)**

**Wigner Function – the Algorithm of the Self (Conscience)**
A neural communication based on **temporal code** whereby different cortical areas which have to contribute to the same percept $P$ **synchronize** their **spikes.** In visual system, spike emission in a single neuron of the higher cortical regions results as **trade-off** between **bottom-up** stimuli arriving through **the lateral geniculate nucleus (LGN)** from the retinal detectors and threshold modulation due to **top-down** signals sent as conjectures by the semantic memory. The „feature binding" hypothesis consists of assuming that all the cortical neurons whose receptive fields are pointing to a specific object synchronize the corresponding spikes. As a consequence the visual cortex organizes into separate neuron groups oscillating on two distinct spike trains for the two objects. Each neuron is a **nonlinear system** passing close to a **saddle point.** While small gradual changes induce the sense of motion **as in movies**, big differences imply completly different subsequent **spike trains.**
Precisely, if we have a sequence of identical spikes of unit area, localized at erratic time positions $\tau_l$ then the whole sequence is represented by: $f(t) = \sum_l \partial(t - \tau_l)$, where $\tau_l$ is the set of position of the spikes. A temporal code, based on the mutual position of successive spikes, depends on the moments of **the interspike interval (ISI)** distributions: $(ISI) = (\tau_l - \tau_{l-1})$. **Different *ISIs* encode different sensory information.** A time ordering **within the sequence** is established by comparing the overlap of two signals as mutually shifted in time. Weighting all shifts with a phase factor and summing up, this amounts to construct a **Wigner function:**
$$W(t, \omega) = \int_{-\infty}^{+\infty} f(t + \frac{\tau}{2}) . f(t - \frac{\tau}{2}) \exp(i\omega\tau) d\tau .$$
If $f$ is the sum of the two packets $f = f_1 + f_2$, the frequency-time plot displays an intermediate interference whatever is the time separation between $f_1$ and $f_2$.
We can model the encoding of external information on a sensory cortical area (V1 in visual case) as a particular spike train assumed by an input neuron directly exposed to the arrival signal and then propagated by coupling through the array. Then transform the local time information into a spatial information which tells the amount of cortical area which is synchronized. If many sites have synchronized by mutual coupling, then the read-out problem consists in tracking the pattern of values, one for each site.

It appears that **the Wigner function** is **the best read-out** („Self – Thinker") of synchronized layer that can be doen by exploiting natural machinery (the right orbitofrontal cortex). The local value of the Wigner function represents a **decision (conscience)** to be sent to a **motor areas** triggering a suitable action. (16) Paradoxically, the free will is involved in quantum measurement **reducing entanglement.**

**Discussion:**

**Self-referential signals** come from motor systems and their sensory components, and are strongly influenced by diffuse ascending value systems (SR= primary consciousness). Other homeostatic systems in the brainstem and periaquaductal gray also contribute to **distinguish self from nonself.** The self emerges and serves as a **reference** (but is not **self-consious**). Only the higher-order consciousness and linguistic capabilities does a **self** arise that is nameable to **itself**. As W. James pointed out, **"the thoughts themselves are the thinker"** = (moving frame of reference in movie of consciousness)**.** (2)

Studies of coupled cultured networks hint that **the neuro-glial fabric** provide a **„photographic film"** like the holograms and that their generation and retrieval is sustained by **chemical waves** generated by **the glia cells** when act as **excitable media.** Our perception seems to be the result of a **sequence of individual snapshots,** a sequence of moments, like individual dicrete **movie frames,** that, when quickly scrolling past us, we experience as **continuos motion.** Events that reach us, are **sequentially** perceived as moving frames in the movie of consciousness at the same moment as synchronous („Carthesian Theater").

As a general rule, we are stuck with the neurons we are born with, but **glia divide and reproduce.** Humans have **higher glia to neuron ratios** (about nine to one) than lower animals. Glia is in continuous dialogue with neuron. Astrocytes are type of glia surrounded the synapse (gap) between neurons. Einstein´s parietal lobe had extremly well developed subcortical white matter. This finding might suggest that Einstein´s **creativity** was related to his enhanced intra- and interhemispheric **connectivity.** The paralell distributed processing (PDP) models also suggests that **information** in PDP networks is stored **in the streghts** of connections between units, as in the brain, and concepts are represented as patterns of activity involving many units, i. e. as distributed representations. (19)

**Area 9** is located in the frontal cortex and is critical in planning behavior, attention, and memory. **Area 39** is located in the parietal lobe and is important for language and spatial functions. By analyzing the number of neurons and glial cells (the supportive tissue of the brain), it was hoped that some histological differences of Einstein´s brain could be uncovered. In this areas **the neuron-to-glia ratio** of Einstein´s brain was much smaller compared to normal brains. The greater number of glial cells per neuron may be indicative of an increased metabolic requirement of Einstein´s brain, his brain needed more energy than of the average man. Thus a hypothesis was formed where the increased energy usage by Einstein´s brain enabled him to perform a **higher level of thinking.**

S. F. Witelson theorized that this unusual parietal and fissure configuration provided more favourable connections between the neurons in this region. Since parietal area is also involved in **math** and **spatial reasoning**, these improved connections could account for Einstein´s mastery of spatial reasoning and advanced mathematics. If it is indeed the unique organization of Einstein´s brain attributed him with such a rare intellect, then there is a specific brain configuration for **genius type individuals ?** (18)

Region near a patch of cortex is called **the TPO** (for the junction of the temporal, parietal, and occipital lobes). The angular gyrus is a part of the TPO that is concerned with numerical concepts such as ordinality (sequence) and cardinality (quantity). In humans visually presented numbers (graphemes) activate cells in the fusiform gyrus, whereas the sounds (**phonemes**) are processed **higher up,** in **the TPO**. Number-color synesthesia might be caused by **cross-wiring** between V4 and the number-appearence area (both within the fusiform) or between the higher color area and the **number-concept area** (both in **the TPO**). Metaphor involves making links between seemingly unrelated conceptual realms. This is not just a **coincidence. The mutation** concerning synesthesia causes **excess communication** among different brain maps (**cross-domain mapping**).

The World as a system can be described due the polar decomposition, as a **composite system** consisting from **two subsystems** (self as subsetA, and non-self as subset nonA) which are mutually observing one another. If there is given some state of one subsystem, then the knowledge of the second system´s state is enable us to know all about the original system, as it coming out from the **envariance.** It means that **the second subsystem has a consciousnsess.** If we are including also **the observer**, we are describing **consciousness**. Cramer´s **objective wave function** can compute the connection between an earlier and later event. **Envariance thinking algorithm** is leading to information without entropy (13), e. i. to **lifting of negentropy** in the system**.**

U. Frith and S. Baron-Cohen posit that the main abnormality in autism is a deficit in the ability **to construct a „theory of other minds". The mirror neurons** are located in the cingulate and insular cortices. They enable humans to **see themselves as others see them**, which is essential ability for **self-awareness.** (20) **Cross-domain mapping** is analogous to metaphors, and involve neural circuits similar to those in the mirror neuron system. Similarities between characteristics of **autism of childs** and **geniality of adults** (A. Einstein) involving a **mirror neuron dysfunction** and **distorted salience landscape**. Above altered limbic connections may led not only to some dysfunctions, but also to **bigger informational capacity** of the brain in other areas. Frequent attentional retreat from the outside world cause decrease in the sensory input („daydreaming", „Sleepwalker"), due a modulation of the synaptic informational transfer lead to increased information capacity in the association areas of the brain. It is not excluding altruistic behavior and empathy (mirror neurons function) as costly signals of general intelligence. Moving frames are perceived as consciousss picture´s flow, i. e. as consciousness in I-mode and the first-person perspective of the thalamocortical system´s FR.

ACKNOWLEDGEMENTS. I am particularly grateful to Vladislav Volman, Vladimir Varga and Konrad Balla for discussions and comments.

**References:**

1. M. Tsang, D. Psaltis: Propagation of temporal entanglement. Physical Review A 73, 013822 (2006)
2. G. M. Edelman: Naturalizing consciousness: A theoretical framework. PNAS.093149100
3. E. Sauser, A. Billard: View Sensitive Cells as a Neural Basis for the Representations of Others in a Self-Centered Frame of Reference. Proceedings of the Third Int. Symposion and Artefacts, Hatfield, U. K. pp. 119-127, (2005)
4. A. Pellionisz: Tensor Geometry: A Language of Brains and Neurocomputers.Generalized Coordinates In Neuroscience and Robotics. Neural Computers, NATO Adv. Res. Workshop, Düsseldorf, Springer Verlag, (1987)

5. T. Wolbers, C. Weiler and C. Büchel: Contralateral Coding of Imagined Body Parts in the Superior Parietal Lobe. Cerebral Cortex, Vol. 13, Number 4, pp. 392-399 (2003)
6. C. Becchio, M. Adenzato, B. G. Bara: How the brain understands intention: Different neural circuits identify the componential features of motor and prior intentions. Consciousness and Cognition 15, pp. 64-74, (2006)
7. G. Cheung, Wai-tian Tan: Loss-Compensated Reference Frame Optimization For Multi-Path Video Streaming. International Packet Video Workshop, Irvine, CA, December 2004
8. G. H. Recanzone: Hearing and Looking. Neuron 29, 314-315, (2001)
9. J. M. Groh, et al. Neuron 29, 509-518, (2001)
10. V. Volman, E. Ben-Jacob, H. Levine: The Astrocyte as a Gatekeeper of Synaptic Information Transfer. Personal communication (2006)
11. F. W. Pfrieger: Role of glia in synapse development. Current Opinion in Neurobiology, 12, pp. 486-490, (2002)
12. W. H. Zurek, R. Blume-Kohout: Quantum Darwinism: Entanglement, Branches, and the Emergent Classicality of Redundantly Stored Quantum Information. ArXiv:quant-ph/0505031 v2 13 Oct (2005)
13. W. H. Zurek: Environment – Assisted Invariance, Causality, and Probabilities in Quantum Physiscs. ArXiv: quant-ph/0211037 v1 7 Nov (2002)
14. M. Schlosshauser: Experimental motivation and empirical consistency in minimal no-collapse quantum mechanics. ArXiv: quant-ph/0506199v3 15 Jan (2006)
15. C. Bedard, H. Kröger, A. Destexhe: Does the $1/f$ frequency-scaling of brain signals reflect self-organized critical states ? arXiv:q-bio.NC/0608026 v1 15 Aug (2006)
16. F. T. Arecchi: Feature Binding as Neuron Synchronization: Quantum Aspects. Brazilian Journal of Physics, vol. 35, no.2A, pp.253-259, (2005)
17. H. Manji in „Glutamate and Disorders of Cognition and Motivation", the New York Academy of Sciences, (2003)
18. M. C. Diamond et all.: On the Brain of a Scientist: Albert Einstein. Experimental Neurology, 88, 198-204, (1985)
19. S. F. Witelson et all.: The Exceptional Brain of Albert Einstein. The Lancet, 353, 2149-2153, (1999)
20. V. S. Ramachandran, L. M. Oberman: Broken Mirrors: A Theory of Autism. Scientific American, October 16, (2006)

**30.11.2006**

**Robert Skopec**
**941 35 Dubník 299**
**Slovakia**

**Mobile: +421 0908220692**